\documentclass[]{aastex631}
\usepackage{multirow}
\usepackage[T1]{fontenc}

\begin{document}

\title{
Upper limits on the gamma-ray emission from the microquasar V4641 Sgr
}
\correspondingauthor{Jian Li}
\email{jianli@ustc.edu.cn}

\author[0000-0003-4936-8300]{Zihao Zhao}
\affiliation{Department of Astronomy, University of Science and Technology of China, Hefei 230026, China}
\affiliation{School of Astronomy and Space Science, University of Science and Technology of China, Hefei 230026, China}
\author[0000-0003-1720-9727]{Jian Li}
\affiliation{Department of Astronomy, University of Science and Technology of China, Hefei 230026, China}
\affiliation{School of Astronomy and Space Science, University of Science and Technology of China, Hefei 230026, China}

\author[0000-0002-1522-9065]{Diego F. Torres}
\affiliation{Institute of Space Sciences (ICE, CSIC), Campus UAB, Carrer de Magrans s/n, E-08193 Barcelona, Spain}
\affiliation{Institut d’Estudis Espacials de Catalunya (IEEC), E-08034 Barcelona, Spain}
\affiliation{Institució Catalana de Recerca i Estudis Avanc˛ats (ICREA), E-08101 Barcelona, Spain}



\begin{abstract}

%
Following a recent detection of TeV radiation 
by the Large High Altitude Air Shower Observatory (LHAASO) and the High-Altitude Water Cherenkov Observatory (HAWC),
coincident with the direction of the microquasar V4641 Sgr, we search for possible GeV emission from this source.
We explored the morphology and temporal features of the source as well as two nearby unassociated point sources which could be a part of extended structure of V4641 Sgr, and compared results with corresponding X-ray and TeV emissions.
%
No significant gamma-ray signal linked to V4641 Sgr 
was detected in the energy range of 1–300 GeV.
The 95\% confidence level upper limits for the flux from the source, assuming both point and extended source models were 5.38$\times$ 10$^{-13}$ erg cm$^{-2}$ s$^{-1}$ and 1.12$\times$ 10$^{-12}$ erg cm$^{-2}$ s$^{-1}$, respectively.
Additionally, no correlation between gamma-ray light curve and X-ray outbursts was observed.
%

\end{abstract}



\section{Introduction} \label{sec:intro}

A microquasar is a binary system that typically includes a compact object, such as a black hole or neutron star, and a main sequence companion star. 
Similarly to quasars, these systems launch relativistic jets, distinguishing them from other populations of X-ray binaries \citep{1999ARA&A..37..409M}. 
In most microquasars, the jet appears in hard/low X-ray states, being heavily suppressed in soft/high states.
Strong flares often occur during transitions from hard to soft states, which are usually linked to X-ray outbursts \citep{2004MNRAS.355.1105F}. 
%
In addition to radio and X-ray emission, microquasars display emission and rapid variability across a broad energy range.
They have been proposed to be potential gamma-ray sources
\citep{1998NewAR..42..579A,1999ARA&A..37..409M}.
In leptonic models, non-thermal relativistic electrons accelerated in the jet are predicted to emit synchrotron gamma radiation and upscatter IR-UV photons to GeV and TeV energy through inverse Compton scattering, see e.g.,  \citet{1998NewAR..42..579A,1999MNRAS.302..253A,2002A&A...388L..25G,2002A&A...385L..10K,2002A&A...393L..61R,2005A&A...429..267B}. 
In a hadronic scenario gamma-ray photons are thought to be produced by the collision between jet protons and ions in the stellar wind from the massive donor or the accretion disk, see e.g., \citet{1992A&A...253L..21M,1996SSRv...75..357A,2003A&A...410L...1R,2005A&A...439..237R,Torres2011}. 
Additionally, the shock developed in jets and in the region where jets impact interstellar medium can also be a source of gamma-ray photons \citep{2011A&A...528A..89B}. 
%
%

Gamma-ray detections of microquasars are still rare, with only a few confirmed cases as SS 433, V4641 Sgr, GRS 1915+105, MAXI J1820+070, Cyg X-1 and Cyg X-3 \citep{2007ApJ...665L..51A,2009Natur.462..620T,2018Natur.562...82A,2024Natur.634..557A,2024arXiv241008988L}.
Recently, the Large High Altitude Air Shower Observatory (LHAASO) collaboration released a catalog of microquasars detected above 25 TeV in its field of view (SS 433, V4641 Sgr, GRS 1915+105, MAXI J1820+070 and Cyg X-1), representing a large improvement of gamma-ray detections of microquasars \citep{2024arXiv241008988L}.
In previous studies, observations of gamma-ray emission from microquasars exhibit different characteristics, including transient, periodic, and steady emission in different energies.
For example, Cyg X-3 shows flare-like GeV signal correlated with spectral state transition of radio and X-ray emission, and the orbital modulation of signal is confirmed \citep{2009Natur.462..620T,2009Sci...326.1512F,2012A&A...545A.110P}.
Cyg X-1 was also found to have several transient VHE (E $\geq$ 100 GeV) events coincident with its X-ray flares 
\citep{2007ApJ...665L..51A}.
V404 Cygni was proposed to have weak episodic GeV flares in its 2015 outburst as well \citep{2016MNRAS.462L.111L,2017ApJ...839...84P,2021ApJ...922..111X}, but the results are controversial \citep{2021MNRAS.506.6029H}.
The gamma-ray flare is always associated with the state transition of microquasar, but origin is still unclear and might be linked to complex process during the launch of relativistic jet.
For steady emission, a point-like GeV signal associated with its hard state was detected for Cyg X-1 near its central engine, which shows a hint of orbital variability \citep{2010ApJ...712L..10S,2013ApJ...766...83S,2016A&A...596A..55Z};
GRS 1915+105 was also claimed to have persistent GeV emission \citep{2024arXiv241010396M};
while periodic GeV emission with the precession period of jets was discovered for SS 433, located at a position of relatively far cloud \citep{2020NatAs...4.1177L}.
%
%
In the case of Cyg X-1, particles are accelerated along jet, and emit GeV photons mostly through inverse-Compton scattering between accelerated particles and seed photons from companion, which results in the observed orbital modulation \citep{2016A&A...596A..55Z}.
For GRS 1915+105, emission could arise from interaction between protons accelerated in jet and nearby gas \citep{2024arXiv241010396M}.
However SS 433's GeV emission may originate from the interaction between relativistic outflow and nearby molecular cloud.
At higher energies, extended and point-like emission was found for several microquasars.
LHAASO detected gamma-ray emission associated with the bow-like radio structure inflated by the jet of Cygnus X-1 above 25 TeV \citep{CygX-1shell,2024arXiv241008988L};
From SS 433, TeV emission from two jet lobes, as well as the central part of the system was detected \citep{2018Natur.562...82A,2024Sci...383..402H,2024arXiv241008988L}, which may have hybrid leptonic and hadronic origin.
GRS 1915+105 and MAXI J1820+070 were found to have extended TeV emission which may be related with their jets \citep{2024arXiv241008988L}.

Recently, the High Energy Stereoscopic System (H.E.S.S.), HAWC and LHAASO discovered TeV extended radiation coincident with the microquasar V4641 Sgr \citep{2024Natur.634..557A,HESS-V4641,2024arXiv241008988L}.
V4641 Sgr is a black hole X-ray binary located $\sim$6 kpc away from earth, consisting of a 6.4 solar mass black hole and a 2.9 solar mass B9III companion \citep{2014ApJ...784....2M}.
The orbital period was measured to be $\sim$2.8 days \citep{2001ApJ...555..489O}.
V4641 Sgr was first discovered in 1999 as an X-ray source. 
Later that year, a radio outburst observed by Very Large Array (VLA) revealed superluminal motion of emitting materials, confirming its nature as a microquasar \citep{2000ApJ...544..977H}. 
The first-observed outburst in 1999 was very bright and short comparing to typical ones of low-mass X-ray binaries (LMXBs) \citep{2022MNRAS.516..124S}.
Thereafter, outbursts of V4641 Sgr were less bright than the one in 1999.
In recent years, V4641 Sgr has exhibited quasi-regular outbursts approximately once every 220 days \citep{2016ApJS..222...15T,2022MNRAS.516..124S}, though few coincident radio flares or state transitions have been detected. 
V4641 Sgr was also found to stay in the soft state with a low X-ray luminosity, which is even lower than typical soft-to-hard transition threshold during outbursts \citep{2010ATel.2832....1Y,2015ATel.7904....1B,2015ApJ...814..158P,2020ATel13443....1A}. 
During the 2015 outburst, Swift/XRT detected that V4641 Sgr transitioned from a soft thermal accretion state to a non-thermal accretion state (essentially equivalent to low/hard state or power law dominated state) when the system should be associated with a compact radio jet, but V4641 Sgr was not detected in radio band either \citep{2015ATel.7908....1M}.
Several V4641 Sgr outbursts occurred during the {\it Fermi}-LAT era, as listed in Table \ref{tab:radio_record}. 
Here, we search for the gamma-ray emission from V4641 Sgr with 15-year accumulated {\it Fermi}-LAT data. 
%

\begin{deluxetable}{lll}[ht!]
\tablehead{\colhead{X-ray outburst date} & \colhead{Radio observation} & \colhead{Reference}}
\tablecaption{X-ray outbursts of V4641 Sgr in the {\it Fermi}-LAT era. \label{tab:radio_record}}
\startdata
    2021 Oct. & no observation & 1 \\
    2020 Jan. & non-detection & 2,3 \\
    2018 Aug. & no observation & 4 \\
    2015 Jul. & non-detection & 5,6 \\
    2014 Jan. & no observation & 7 \\
    2010 Jun. & non-detection & 8,9 \\
\enddata
\tablerefs{1: \citet{2021ATel14968....1N}; 2: \citet{2020ATel13431....1S}; 3: \citet{2020ATel13471....1Z}; 4: \citet{2018ATel11931....1N}; 5: \citet{2015ATel.7858....1Y}; 6: \citet{2015ATel.7908....1M}; 7: \citet{2014ATel.5803....1T}; 8: \citet{2010ATel.2785....1Y}; 9: \citet{2010ATel.2832....1Y}}
\end{deluxetable}


\section{Observations and data analysis} \label{sec:data}

We used 15 years of {\it Fermi}-LAT data between Aug 4th, 2008 and Sep 21th, 2023, to search for the gamma-ray emission of V4641 Sgr. 
All events from ``P8 Source" event class (corresponding to ``evclass=128" and ``evtype=3") in an energy range of 1 GeV-300 GeV was selected,  within 15 degrees of V4641 Sgr's optical position (ra=274.840, dec=-25.407) \citep{2020yCat.1350....0G}. 
The energy threshold of 1 GeV was applied to take advantage of better PSF and less galactic diffuse emission at higher energies \citep{2024arXiv241010396M}.
We used the instrument response function ``P8R3\_SOURCE\_V3". 
%
The zenith angle was limited to 90 degrees to reduce contamination from Earth's limb.
We built the source model from the Fermi-LAT 4FGL-DR4 catalogue \citep{2023arXiv230712546B} to account for background source contributions. 
All sources within 20 degrees of V4641 Sgr have been included. 
Since V4641 Sgr locates in a crowded region close to the galactic plane, only the spectral parameters of sources within 4 degrees of V4641 Sgr were set to free while all other sources were fixed to their catalog values.
The galactic and isotropic diffuse emission components, ``gll\_iem\_v07" and ``iso\_P8R3\_SOURCE\_V3\_v1", were included in the background model. 
\textit{Fermitools 2.2.0} and \textit{Fermipy 1.2.0} were used for the analysis.

\begin{figure}[hb]
    \plotone{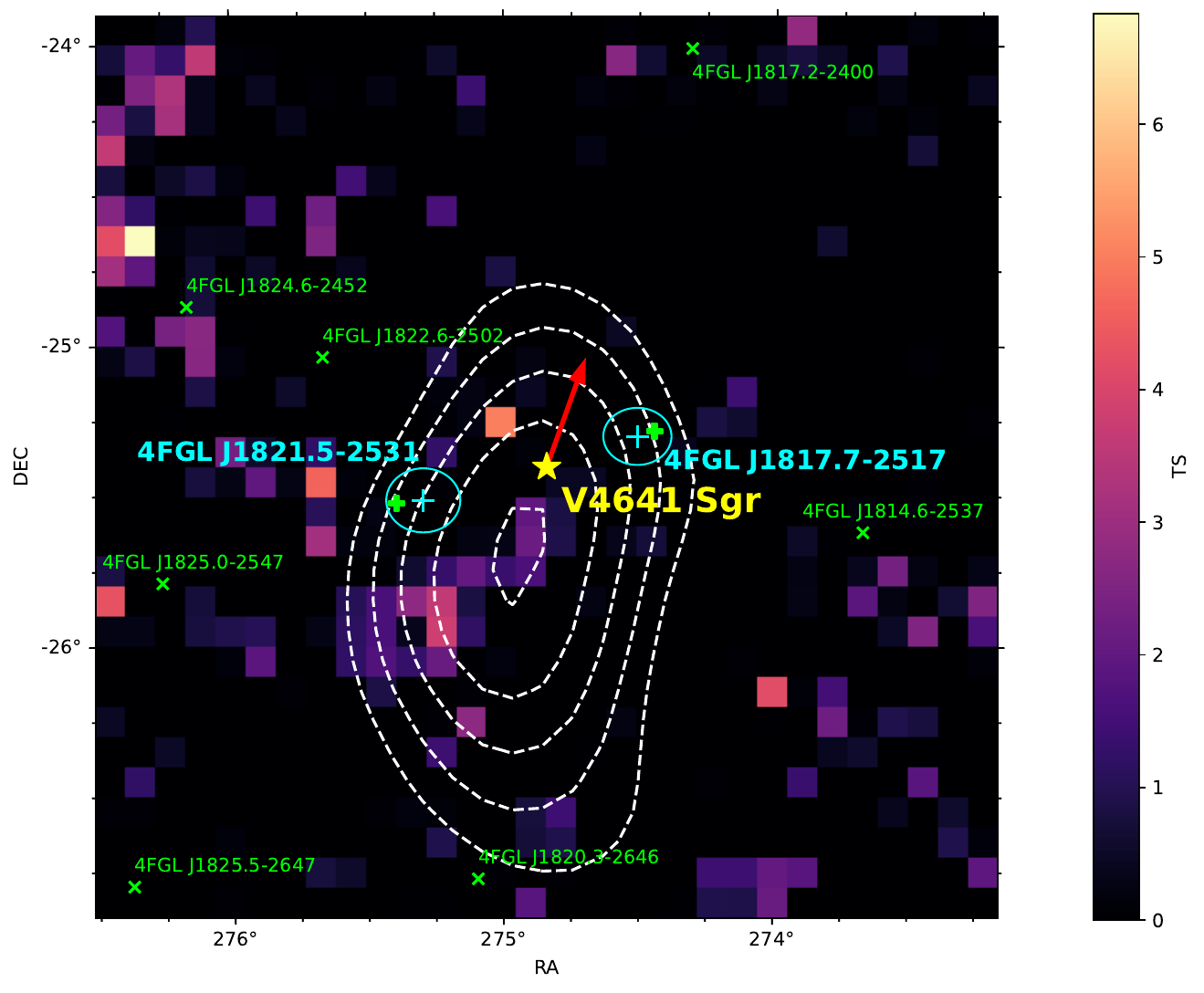}
    \caption{The TS map of a $3^{\circ} \times 3^{\circ} $ field around V4641 Sgr. 
    The background sources from 4FGL are shown with green crosses.
    The updated position of 4FGL J1821.5-2531 and 4FGL J1817.7-2517 are shown in cyan with their 99\% uncertainty ellipse, while their 4FGL position are shown with green filled plus.
    The position of V4641 Sgr is shown with a yellow star.
    The white dashed lines show the TeV contour observed by LHAASO, starting from 4$\sigma$ with a step of 1$\sigma$.
    Red arrow represents the direction of apparent superluminal jet ejecta in V4641 Sgr's 1999 outburst \citep{2000ApJ...544..977H}.
    }
    \label{fig:resid_tsmap}
\end{figure}

\section{Results} \label{sec:result}

\subsection{Morphology and Spectrum Analysis} \label{subsec:mp}

\begin{figure}[ht!]
    \plotone{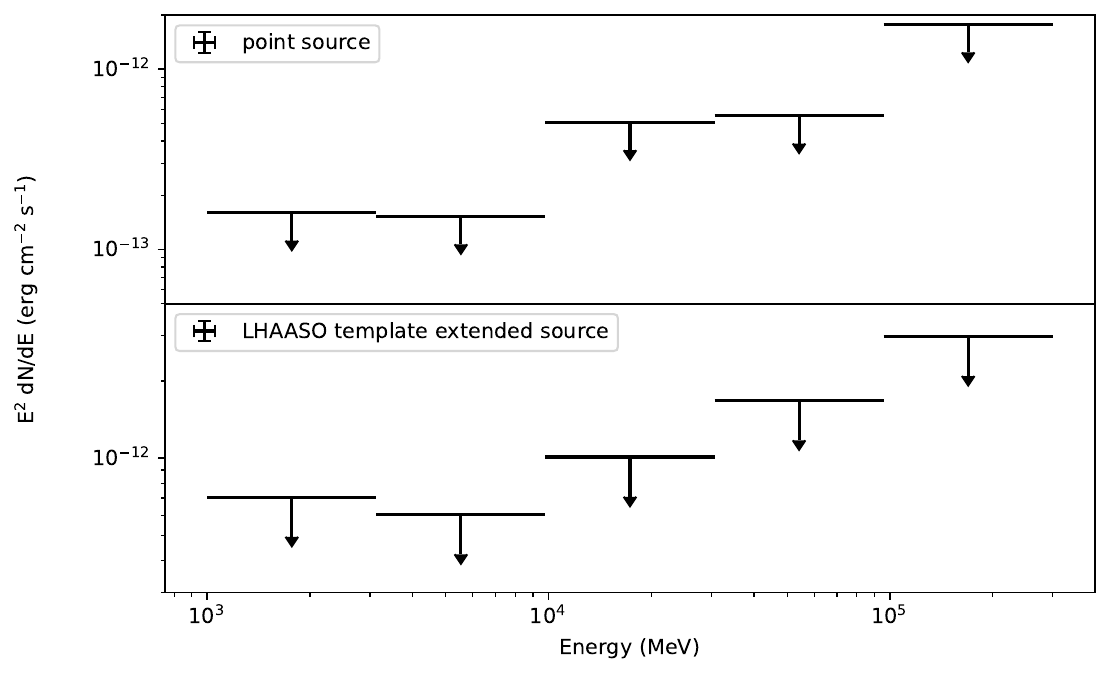}
    \caption{The SEDs of V4641 Sgr as a point source (top) and extended source with the LHAASO template (bottom). 
    95\% flux upper limits imply that the TS value is less than 4.}
    \label{fig:SED}
\end{figure}

There is no known 4FGL-DR4 source associated with V4641 Sgr.
However, two nearby unidentified point sources were noted: 4FGL J1821.5-2531 and 4FGL J1817.7-2517. 
Their positions have been updated as following, 4FGL J1821.5-2531: RA=$275.30 \pm 0.04
^{\circ}$, DEC=$-25.52 \pm 0.04^{\circ}$; 4FGL J1817.7-2517: RA=$274.51 \pm 0.04^{\circ}$, DEC=$-25.31 \pm 0.03^{\circ}$. 
We conducted binned likelihood analysis to explore the significance of V4641 Sgr.
The residual TS map of this region (Figure \ref{fig:resid_tsmap}) showed no significant TS excess (TS $>$ 25) at the position of V4641 Sgr.
By adding a point source at the position of V4641 Sgr, we derived a TS value of 0.1 and a flux upper limit of 5.38$\times$ 10$^{-13}$ erg cm$^{-2}$ s$^{-1}$ at 95\% confidence level in the 1-300 GeV band.
We tested the possibility of the existence of a weak extended source around V4641 Sgr. 
The radial Gaussian and radial Disk models were adopted, but no detection was found. 
We further adopted the LHAASO significance map of V4641 Sgr \citep{2024arXiv241008988L} as a template (Figure \ref{fig:resid_tsmap}), leading to a TS value of 0 and a flux upper limit of 1.12$\times$ 10$^{-12}$ erg cm$^{-2}$ s$^{-1}$ at 95\% confidence level in 1-300 GeV.
The corresponding flux upper limits of V4641 Sgr at different energies are shown in Figure \ref{fig:SED}.
The extension of 4FGL J1821.5-2531 and 4FGL J1817.7-2517 have also been explored, but no significant extension was detected. 
TS of their best-fit extension are 0.4 and 0 respectively.
We further tested the model of one extended source.
The best-fit Radial Disk extended model is at the position of RA = 275.2 $\pm$ 0.9$^{\circ}$, DEC = -25.5 $\pm$ 0.4$^{\circ}$, with a radius of 0.31 $\pm$ 0.06$^{\circ}$. 
Comparing to two point-source model, the $\Delta$ TS is -48 and the $\Delta$ AIC (Akaike Information Criterion \citep{1974ITAC...19..716A}) is 40.
The best-fit Radial Gaussian extended model locates at RA = 274.6 $\pm$ 0.1$^{\circ}$ ,DEC = -25.3 $\pm$ 0.1$^{\circ}$, with a sigma of 0.31 $\pm$ 0.09$^{\circ}$.
The $\Delta$ TS is -58 and the $\Delta$ AIC is 50 compared to two-point-sources model.
Therefore, this region is better described by two point sources.

\begin{figure*}[htbp]
\includegraphics[width=\textwidth]{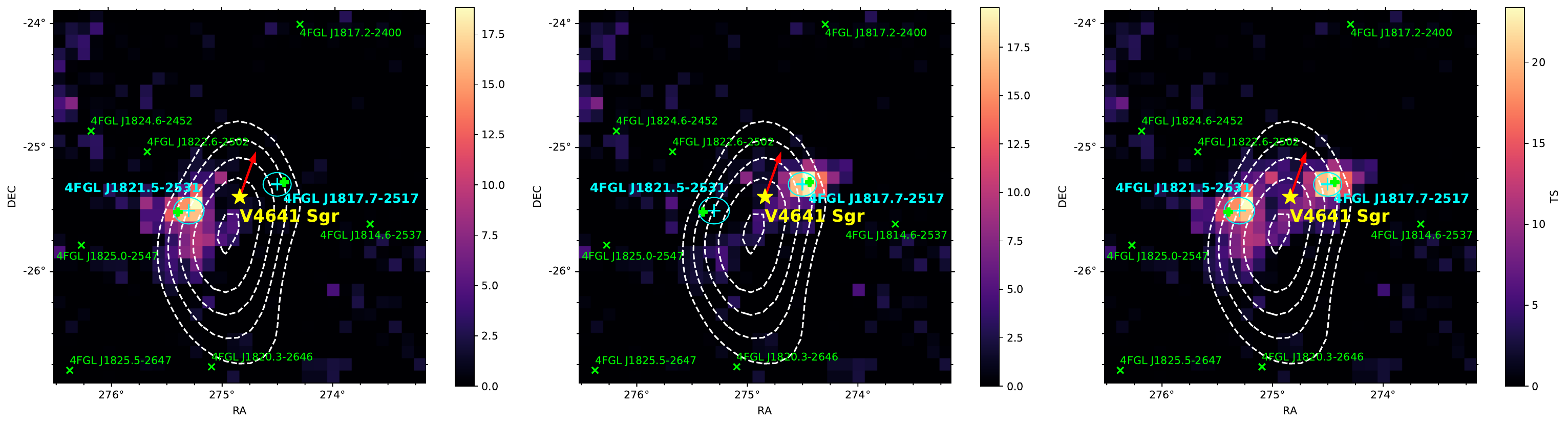}
    \caption{TS map for 4FGL J1821.5-2531 (left), 4FGL J1817.7-2517 (middle) and combined (right). Labels and contours are the same as in Figure \ref{fig:resid_tsmap}.}
    \label{fig:model_tsmap}
\end{figure*}

Since no gamma-ray emission is detected from the position of V4641 Sgr, we explored whether the two nearby point sources 4FGL J1821.5-2531 and 4FGL J1817.7-2517, which locates on either side of V4641 Sgr, are related to V4641 Sgr.
The TS values of 4FGL J1821.5-2531 and 4FGL J1817.7-2517 in 1-300 GeV are 28.9 and 46.1, respectively, and they locate at the proximity of the very high energy gamma-ray emission observed by LHAASO (Figure \ref{fig:resid_tsmap}). 
Figure \ref{fig:model_tsmap} shows TS maps of the two sources.
Based on the study of broad emission feature above 4 keV in the X-ray spectrum, \citet{2014MNRAS.438L..41G} found that the possible X-ray jet shows a low inferred inclination angle $i_{jet} < 10^{\circ}$, which is consistent with the apparent superluminal jet ejecta in 1999 \citep{2000ApJ...544..977H,2001ApJ...555..489O}.
This direction (as shown in Figure \ref{fig:model_tsmap}) does not align with the two sources.
Such spatial discrepancy make it difficult to explain 4FGL J1821.5-2531 and 4FGL J1817.7-2517 as V4641 Sgr's jet lobes.

\begin{figure}[htbp]
\includegraphics[width=\textwidth]{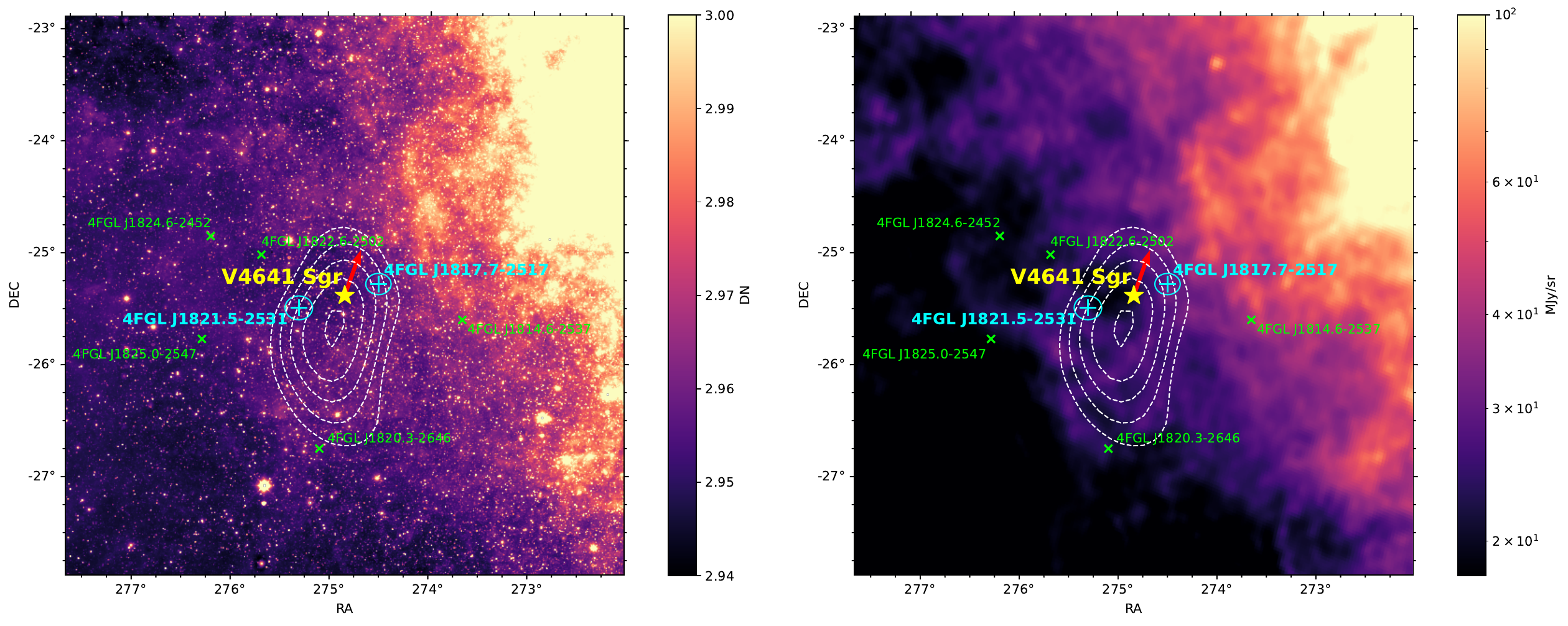}
    \caption{Left: WISE 12 $\mu m$ infrared emission map. Right: Planck dust map. Labels and contours are the same as those shown in Figure \ref{fig:resid_tsmap}.}
    \label{fig:clouds_map}
\end{figure}

Observations on microquasars and ultra luminous X-ray sources (e.g. Cyg X-1 \citep{2005Natur.436..819G,2007MNRAS.376.1341R} and the NGC 55 ULX-1 \citep{2023ApJ...955...61Z}) revealed that the relativistic jet and outflow can produce bubble-like or shell-like structures around the central object when they interact with the surrounding medium. 
Figure \ref{fig:clouds_map} shows sky maps of the infrared emission measured by WISE \citep{2010AJ....140.1868W} and the Planck dust distribution \citep{2020A&A...641A..12P}.
No apparent cavity or shell-like structure could be found around V4641 Sgr. 
Adopting the HI4PI data \citep{2016A&A...594A.116H}, we observed two velocity components in the HI spectra, peaking at 6.54 km/s and 19.42 km/s, respectively.
The 6.54 km/s component corresponds to a distance of $2.41 \pm 2.94$ kpc while the 19.42 km/s corresponds to a distance of $4.19 \pm 2.02$ kpc \citep{2016ApJ...823...77R,2019ApJ...885..131R}. 
Neither matches the distance of V4641 Sgr.

\subsection{Variability Analysis} \label{subsec:lc}

\begin{figure}[!h]
    \includegraphics[width=0.9\linewidth]{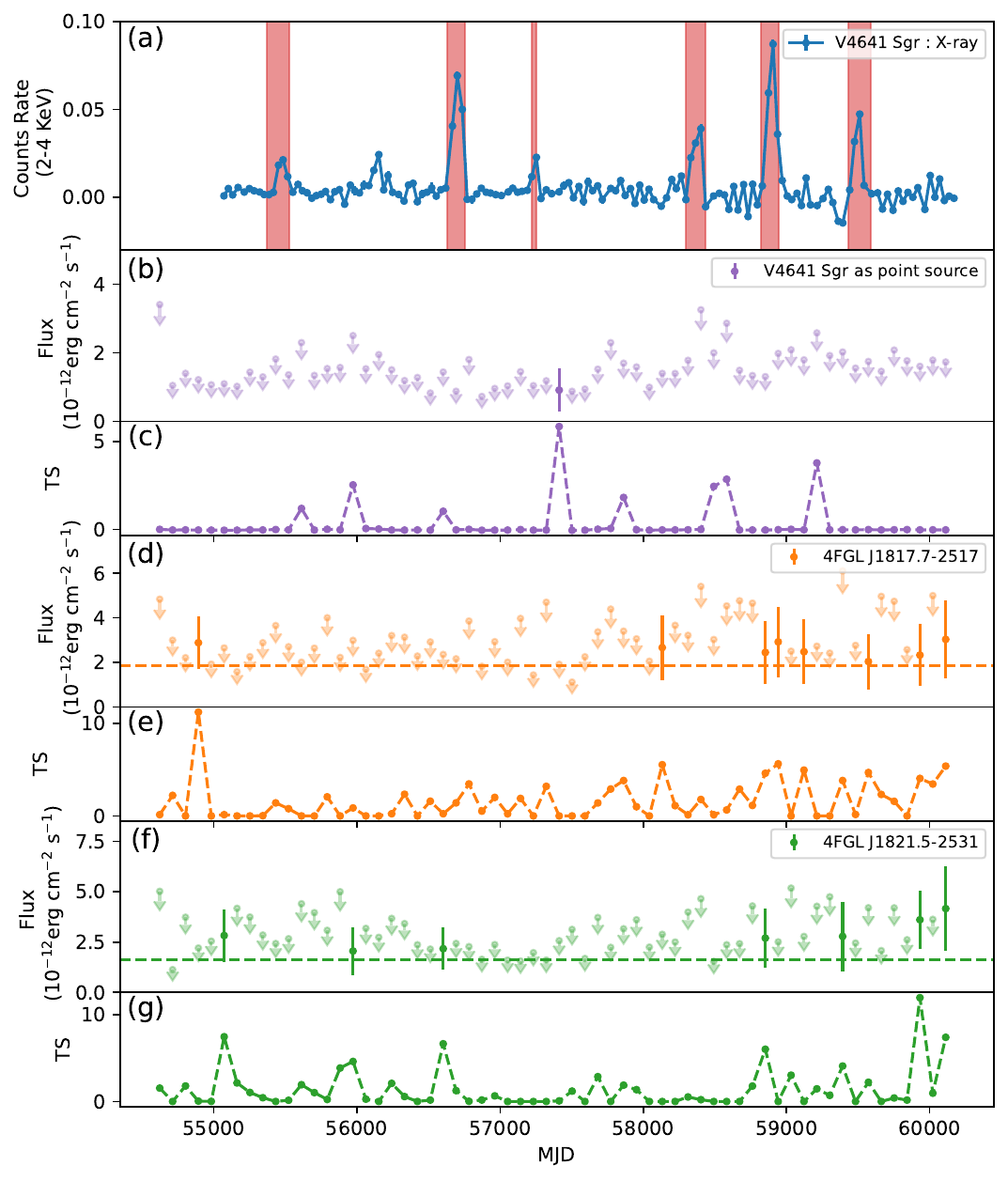}
    \caption{
    Panel (a): 2-4 keV X-ray light curve of V4641 Sgr.
    Panel (b) and (c): 1-300 GeV light curves (flux $\&$ TS value) of V4641 Sgr (assuming a point source).
    Panel (d) and (e): 1-300 GeV light curves (flux $\&$ TS value) of 4FGL J1817.7-2517.
    Panel (f) and (g): 1-300 GeV light curves (flux $\&$ TS value) of 4FGL J1821.5-2531.
    95\% flux upper limits were calculated when TS is less than 4. 
    The last bin of light curve has been removed because of much shorter observation time (6 days) than 90 days. 
    Dashed horizontal line in panel (d) and (f) represent the average flux level of 4FGL J1817.7-2517 and 4FGL J1821.5-2531 derived from 15-year dataset analysis.}
    \label{fig:likelihood_lc}
\end{figure}

We searched for potential correlated variability of V4641 Sgr between X-rays and gamma rays.
Assuming a point source with power-law spectrum, we produced a gamma-ray light curve in 1--300 GeV, via binned likelihood analysis with a binning of 90 days. 
Only the normalization parameter of sources within 4 degrees of V4641 Sgr are allowed to vary.
For our sources of interest, their spectral index are fixed to best fitted values using 15 years data.
The 2--20 keV X-ray light curve, with a binning of 30 days, was obtained from the MAXI mission \citep{2009PASJ...61..999M}.
X-ray and gamma-ray light curves are shown in Figure \ref{fig:likelihood_lc} for comparison. 
No correlated activity was observed and V4641 Sgr stayed below the detection threshold through the whole observation period.
The gamma-ray light curves of 4FGL J1821.5-2531 and 4FGL J1817.7-2517 (Figure \ref{fig:likelihood_lc}) have also been produced and no correlation with X-ray activity could be observed.
We further adopted Z-transformed Discrete Correlation Function (ZDCF) \citep{1997ASSL..218..163A} to evaluate the correlation between X-ray flux light curve (panel (a) in Figure \ref{fig:likelihood_lc}) and TS light curves (panel (c), (e) and (g) in Figure \ref{fig:likelihood_lc}).
It allows us to estimate the cross-correlation of sparse and unevenly sampled astronomical time-series.
We found no significant correlation (> 3$\sigma$) between X-ray light curve and TS light curves of V4641 Sgr and the two nearby point sources (4FGL J1821.5-2531 and 4FGL J1817.7-2517), for any time lag shorter than 200 days.
The most significant correlation between X-ray light curve and TS light curves of V4641 Sgr, 4FGL J1821.5-2531 and 4FGL J1817.7-2517 is 1.0$\sigma$, 2.3$\sigma$ and 2.8$\sigma$ with time lag of -171 days, 28 days and 148 days, respectively.

\begin{figure}[ht]
    \includegraphics[width=\textwidth]{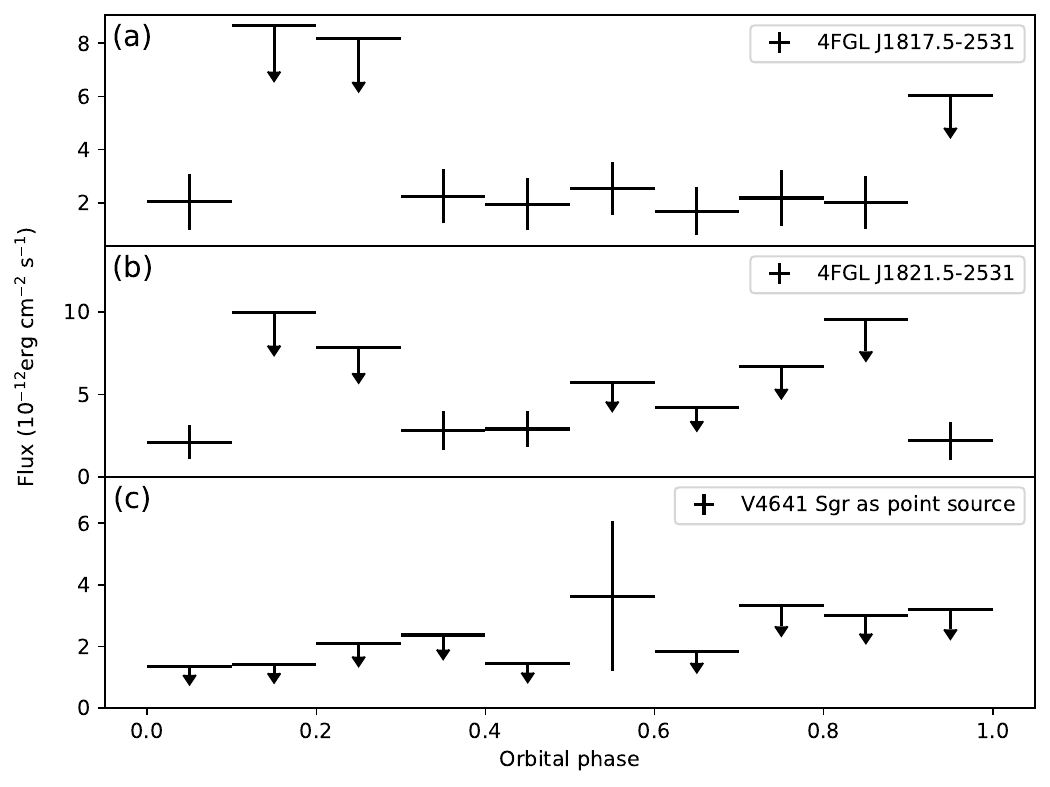}
    \caption{Panel (a) (b) and (c): the orbital phase light curve of  4FGL J1817.7-2517, 4FGL J1821.5-2531, and V4641 Sgr (assuming a point source), respectively. 
    The binsize is 0.1. 95\% flux upper limit was calculated when TS is less than 4.}
    \label{fig:orb_lc}
\end{figure}

We also generated orbital light curves for V4641 Sgr, 4FGL J1821.5-2531, and 4FGL J1817.7-2517, adopting the orbital period of $P_{orb}= 2.81730 \pm 0.00001$ days and epoch of superior conjunction at $T_0=2447707.4865 \pm 0.0038 $ HJD \citep{2001ApJ...555..489O}. 
The corresponding orbital light curves are shown in Figure \ref{fig:orb_lc} and no significant orbital modulation is observed.

Finally, we carried out a search for gamma-ray emission during the X-ray outburst periods (Table \ref{tab:radio_record}; the time range marked with red shading in Figure \ref{fig:likelihood_lc}), separating it from quiescence. 
During X-ray outbursts, the TS of V4641 Sgr, 4FGL J1817.7-2517, and 4FGL J1821.5-2531 are 0, 9 and 2.7, respectively, indicating none of them has experienced a significant brightening. 
In the quiescence period, instead, TS of V4641 Sgr, 4FGL J1817.7-2517 and 4FGL J1821.5-2531 are 0, 37.8 and 24.9, respectively.
%
%
Thus, neither during the X-ray outbursts nor in quiescent was there any significant gamma-ray excess from V4641 Sgr.

\section{Discussion and Conclusion} \label{sec:discussion}

In summary, we searched for gamma-ray emission from V4641 Sgr and none was found.
Both the morphology and variability analyses provided no evidence for a correlation between V4641 Sgr and the two point sources nearby, 4FGL J1817.7-2517 and 4FGL J1821.5-2531. 
We explored the scenario that a jet or an outflow from V4641 Sgr interacts with a dense environmental medium, but no dense gas cluster or bubble structure was found in this area. 
V4641 Sgr as well as the two nearby point sources, 4FGL J1817.7-2517 and 4FGL J1821.5-2531, shows no correlation with the X-ray outbursts of V4641 Sgr. 
Finally, no orbital modulation was observed from V4641 Sgr, 4FGL J1817.7-2517 and 4FGL J1821.5-2531. 
Thus, we conclude that no GeV gamma-ray emission from V4641 Sgr is detected in {\it Fermi}-LAT observations.
4FGL J1817.7-2517 and 4FGL J1821.5-2531 are independent gamma-ray sources rather than extended structure of V4641 Sgr.
An AI-based study classified 4FGL J1817.7-2517 and 4FGL J1821.5-2531 as pulsar and flat spectrum radio quasar, respectively, according to parameters in 4FGL catalog \citep{2021MNRAS.505.5853G}.
But no actual counterparts have been identified in observations yet.

%
%
%
Gamma-ray emission from microquasars is thought to be tied to a jet, where particles are supposed to be accelerated up to high energy, see, e.g., \citet{2009IJMPD..18..347B}.
However, V4641 Sgr has not shown evidence for the relativistic radio jet for a long time span 
(Table \ref{tab:radio_record}).
There is no record of a clear state transition or an intermediate state after 2008, except the soft-to-hard transition at the end of 2015 outburst \citep{2015ATel.7966....1S}.
%
Instead, different to other black hole X-ray binaries, V4641 Sgr remains in soft state even at a very low luminosity \citep{2015ATel.7904....1B,2015ApJ...814..158P}. This may have suppressed the launching of transient relativistic jets, leading to their absence in radio observations post-2008.
However, despite of the lack of radio observations on V4641 Sgr in hard/low state, the low X-ray Eddington ratio $-7 < \log{(L_{X} / L_{Edd})} < -4 $ \citep{2014MNRAS.438L..41G,2015ApJ...814..158P} could still lead to a weak steady jet capable of particle acceleration and gamma-ray emission (e.g. as in Cyg X-1 \citep{2016A&A...596A..55Z}).
%
%
%
%
%
%
%
\citet{2024arXiv241017608N} proposed a hadronic origin of V4641 Sgr's gamma-ray emission, as a result of high-energy cosmic-ray particles escaping from the microquasar along the Galactic magnetic field lines and interacting with the interstellar medium.
Flux upper limits of V4641 Sgr were also calculated by \citet{2024arXiv241017608N} using a photometry method.
Their prediction is consistent with the upper limits of V4641 Sgr as an extended source with the LHAASO template presented in this paper, which are derived with complete likelihood analysis and thus more accurate.

On the other hand, the central engine of V4641 Sgr is claimed to be heavily obscured, possibly by an optically thick equatorial outflow or a large scale-height accretion flow \citep{2020A&A...639A..13K,2022MNRAS.516..124S}.
In this scenario, the photon-photon absorption could play an important role near the compact object.
The UV and soft X-ray photons in the surrounding plasma may interact with GeV gamma rays generated in inner region, causing strong absorption \citep{2009IJMPD..18..347B} and leading to a high GeV optical depth,
%
%
while the TeV emission region may locate at a larger distance from the compact object judging from its morphology, suffering less absorption.
\begin{acknowledgments}
This work is supported by National Natural Science Foundation of China (NSFC) Programs No.12273038.
DFT acknowledges support from the Spanish grants PID2021-124581OB-I00, 2021SGR00426, CEX2020-001058-M and EU PRTR-C17.I1.
The \textit{Fermi} LAT Collaboration acknowledges generous ongoing support
from a number of agencies and institutes that have supported both the
development and the operation of the LAT as well as scientific data analysis.
These include the National Aeronautics and Space Administration and the
Department of Energy in the United States, the Commissariat \`a l'Energie Atomique
and the Centre National de la Recherche Scientifique / Institut National de Physique
Nucl\'eaire et de Physique des Particules in France, the Agenzia Spaziale Italiana
and the Istituto Nazionale di Fisica Nucleare in Italy, the Ministry of Education,
Culture, Sports, Science and Technology (MEXT), High Energy Accelerator Research
Organization (KEK) and Japan Aerospace Exploration Agency (JAXA) in Japan, and
the K.~A.~Wallenberg Foundation, the Swedish Research Council and the
Swedish National Space Board in Sweden.
Additional support for science analysis during the operations phase is gratefully
acknowledged from the Istituto Nazionale di Astrofisica in Italy and the Centre
National d'\'Etudes Spatiales in France. This work performed in part under DOE
Contract DE-AC02-76SF00515.

\end{acknowledgments}

%

\vspace{5mm}
\facilities{{\it Fermi}-LAT, MAXI, WISE, Planck}


\software{Fermitools, Fermipy}




\bibliography{v4641}{}
\bibliographystyle{aasjournal}



\end{document}